\documentclass[sigconf]{acmart}

\usepackage{booktabs} 
\usepackage{enumitem}
\usepackage{tikz}
\usepackage{caption}
\usetikzlibrary{matrix,chains,positioning,decorations.pathreplacing,arrows}
\usepackage{wpmacros}  
\usepackage{url}
\usepackage{multirow}
\hypersetup{draft}

\copyrightyear{2017}
\acmYear{2017}
\setcopyright{acmcopyright}
\acmConference{DLRS 2017}{August 27, 2017}{Como, Italy}\acmPrice{15.00}\acmDOI{10.1145/3125486.3125496} \acmISBN{978-1-4503-5353-3/17/08}

\begin{document}	

\title{Auto-Encoding User Ratings via Knowledge Graphs in Recommendation Scenarios}

\author{Vito Bellini}
\affiliation{%
	\institution{Polytechnic University of Bari}
	\streetaddress{Via E. Orabona, 4}
	\city{Bari} 
	\state{Italy} 
	\postcode{70126}
}
\email{vito.bellini@poliba.it}

\author{Vito Walter Anelli}
\affiliation{%
	\institution{Polytechnic University of Bari}
	\streetaddress{Via E. Orabona, 4}
	\city{Bari} 
	\state{Italy} 
	\postcode{70126}
}
\email{vitowalter.anelli@poliba.it}

\author{Tommaso Di Noia}
\affiliation{%
  \institution{Polytechnic University of Bari}
  \streetaddress{Via E. Orabona, 4}
  \city{Bari} 
  \state{Italy} 
  \postcode{70126}
}
\email{tommaso.dinoia@poliba.it}

\author{Eugenio Di Sciascio}
\affiliation{%
	\institution{Polytechnic University of Bari}
	\streetaddress{Via E. Orabona, 4}
	\city{Bari} 
	\state{Italy} 
	\postcode{70126}
}
\email{eugenio.disciascio@poliba.it}

\renewcommand{\shortauthors}{V. Bellini et al.}

\begin{abstract}
In the last decade, driven also by the availability of an unprecedented computational power and storage capabilities in cloud environments, we assisted to the proliferation of new algorithms, methods, and approaches in two areas of artificial intelligence: knowledge representation and machine learning. On the one side, the generation of a high rate of structured data on the Web led to the creation and publication of the so-called knowledge graphs. On the other side, deep learning emerged as one of the most promising approaches in the generation and training of models that can be applied to a wide variety of application fields. More recently, autoencoders have proven their strength in various scenarios, playing a fundamental role in unsupervised learning. In this paper, we instigate how to exploit the semantic information encoded in a knowledge graph to build connections between units in a Neural Network, thus leading to a new method, SEM-AUTO, to extract and weight semantic features that can eventually be used to build a recommender system. As adding content-based side information may mitigate the cold user problems, we tested how our approach behaves in the presence of a few ratings from a user on the Movielens 1M dataset and compare results with BPRSLIM.
\end{abstract}

%
%


\keywords{Recommender Systems, Deep Learning, Autoencoders, Knowledge graphs, Linked Open Data, DBpedia}

\maketitle

\section{Introduction}\label{sec:intro}
Recommender systems (RS) are nowadays used in many of the services we daily access in order to provide a personalized experience in the browsing and selection of items in a catalogue. Their success strongly depends on how they can identify and exploit user tastes and preferences while suggesting potentially relevant items. RS mainly rely on the rates users provide to items, to predict the importance for unrates ones. Over the years, collaborative filtering (CF) approaches have shown to be very effective in suggesting accurate recommendations especially in the presence of many data in the user-ratings matrix while they suffer in situations of very sparse matrices. CF fail in providing good recommendation in situations where we have users who rated a few items (cold users) and items with a few ratings (cold items). This latter problem is mitigated in case the recommendation engine adopts a content-based (CB) approach where  characteristics of the items are exploited to find those similar to the ones rated by the user in the past. Combining CF and CB usually leads to obtain better recommendation results \cite{Khrouf2013}\cite{Meymandpour:15}\cite{Ostuni2013}. 
As a matter of fact, adding side information to a collaborative filtering approach has proven to be more effective while computing recommendations to the end user \cite{Ning12}. Recently, among the ideal candidates to get side information to be injected in recommender systems we surely find knowledge graphs\footnote{\url{https://googleblog.blogspot.it/2012/05/introducing-knowledge-graph-things-not.html}}. On the one hand, information encoded in such structures is an excellent mine of meaningful data that can be exploited to describe and categorize items in a catalogue. Among the various knowledge graphs available online, we have those belonging to the Linked Open Data (LOD) cloud such as DBpedia \cite{dbpedia2007} or Wikidata \cite{wikidata2014}. There, encyclopedic information is encoded in terms of RDF triples $\langle subject, predicate, object \rangle$ thus creating a huge interconnected graph of knowledge. 

After being successfully adopted to cope with many tasks related to artificial intelligence such as image recognition or natural language processing, deep learning techniques are rapidly entering the world of recommender systems  \cite{covington2016deep}. Built around the basic notion of a neural networks (NN), over the years many new techniques and approaches have been developed under the deep learning umbrella. Among them, autoencoders are a particular configuration of a NN whose main aim is that of training a model able to reproduce the inputs of a system. They have been successfully adopted for feature selection \cite{vincent2008extracting} as well as a generative model of data \cite{kingma2013auto}.

In this paper we present a novel way to build a user profile by using features computed by means of an Autoencoder. The semantic information of classes and categories of a knowledge graph is exploited to draw the topology of the underlying neural network. Each class and category associated to an item is represented by a neuron in the hidden layer of the network which autoencodes the ratings of the users. After training the resulting neural network, the weights computed as inputs of the neurons in the hidden layer are then interpreted as the importance of the corresponding feature in the rating process from the users. Eventually, the vectors of feature weights are used to estimate the utility associated to items unknown to (unrated by) the user thus computing a top-N recommendation list.

The remainder of this paper is structured as follows: in the next section we report related work on the usage of autoencoders and deep learning techniques as well as Linked Open Data for recommendation tasks. Then we introduce the notion of autoencoder and that of semantic-autoencoder. In Section \ref{sec:recommendations} we describe our recommendation model followed by the description of the experimental setting and of the metrics we used in the evaluation. Conclusions and Future Work close the paper.

\section{Related work}\label{sec:related}
\noindent \textbf{Autoencoders and Deep Larning. } Autoencoders have recently attracted attention in the Recommender System community. In \cite{Wu2016} the authors utilize the idea of Denoising Auto-Encoders for learning from corrupted inputs. The proposed approach assumes that observed user-item interactions are a corrupted version of user's preferences. The model then learns a latent representation of corrupted user-item preferences that can lead to a better reconstructed input. By training on corrupted data we can recover co-preference patterns. The authors show that this is an effective approach in collaborative filtering scenarios. \cite{Strub:2016:HRS:2988450.2988456} introduces a CF approach based on Stacked Denoising Autoencoders in order to learn a non-linear representation of the users-items in order to alleviate the cold start problem by integrating side information. So they came up with a Hybrid Recommender System. The main idea of this work is to use a hidden layer (Autoencoder's bottleneck) of size k << N (number of features) to let the network find the low-dimensional representation to feed a Deep Neural Network. Experimental results show that side information brings only a small improvement if an item has many ratings. A pure CF approach has been settled with autoencoders in \cite{Sedhain:2015:AAM:2740908.2742726}. They compare item-based autoencoding and user-based autoencoding, outperforming all the baselines in terms of RMSE. A Hybrid Recommender System is presented in \cite{AAAI1714676}. Here the authors use side information to address the problem of the sparse user-item matrix, then jointly learn users and items' latent factors from side information and collaborative filtering from the rating matrix. Other works like \cite{Vuurens:2016:EDS:2988450.2988457} are focused on learning users preferences in a high-dimensional semantic latent space, with the advantage of being able to recommend items using content that describes the items. As the authors say, describing items in a semantic space provides the intersubstitutability of items or, in other words, items may be substituted by nearby items in such a space. Deep Learning is used in \cite{Wang:2015:CDL:2783258.2783273} to tackle CF's sparsity problem by integrating auxiliary information, such as item content information. Ratings and side information can be used together, where a collaborative topic regression method is capable to learn a latent representation. Then, a collaborative deep learning model jointly performs representation for content information and collaborative filtering for the ratings matrix. Mapping user and items to a latent space, as done in \cite{Elkahky:2015:MDL:2736277.2741667} seems to be a good approach to address the recommendation quality in content-based recommender systems. Using a rich feature set from different domains to represent users and items, allows the model to provide quality recommendation across all domains, as well as having a semantically rich user latent feature vector.
\noindent \textbf{Linked Open Data. } Many approaches have been proposed for exploiting information extracted from Linked Open Data in recommendation tasks. One of the very first proposals in this direction is \cite{Heitmann:2010:firstLODrecsys} where the authors introduced for the first time the idea of using Linked Open Data in a recommender system. A system for recommending artists and music using DBpedia was presented in \cite{Passant2010}. Several other approaches have been proposed afterwards such as a knowledge-based framework leveraging DBpedia for the cross-domain recommendation task \cite{Fernandez-Tobias2011,tomeoetal2016}, a content-based context-aware method able to adopt a semantic representation based on a combination of distributional semantics and entity linking techniques \cite{MustoSLG14}, a hybrid graph-based algorithm based on learning-to-rank method and path-based features extracted from heterogeneous information networks built upon DBpedia and collaborative information \cite{DiNoia:2016:SPrank}. To the best of our knowledge, the only works dealing with automated feature selection from knowledge graphs are \cite{Musto-UMAP16-LOD,Ragone-2017}. While the former analyzes the performance of a recommender system after a feature selection based on classical statistical methods such as Information Gain, Chi Squared etc., the latter adopts a technique based on ontological schema summarization.

\section{Autoencoders}\label{sec:autoencoders}
Artificial Neural Networks (ANNs) are computational models originally proposed to catch underlying relationships in a set of data by using positive and negative examples fed into the network (supervised learning).
ANNs are composed by an input layer, one o more hidden layers and an output layer. Every layer is made of units and every layer is fully-connected, meaning that every unit is connected with all the units in the following layer. Connections between units are usually initialized with a random weight. In a conventional neural network, the task is then predicting a target vector $y$ from an input vector $x$.

Autoencoders are ANNs that apply the backpropagation algorithm, setting the target values to be equal to the inputs in an unsupervised fashion. 
Roughly, in an autoencoder network one tries to ``predict'' $x$ from $x$. The idea is to  first compress (encode) the input vector to fit in a smaller representation, and then try to reconstruct (decode) it back. This means that the model learns in the hidden layers, a representation of the input and therefore a latent representation of the knowledge behind the input data.
The task performed by autoencoders is quite similar to that of a Principal Component Analysis (PCA) operation. We suppose to have a two layers ANN, with only one hidden layer using linear functions, the number of hidden units is less than the number of input units, the input and output layers composed by the same number of neurons, and we want the output to mimic as close as possible the input.
After training, the hidden layer will be a representation of the input in a space with a number of dimensions denoted by the number of neurons of the hidden layer. It is noteworthy that the overall training is proportional to the number of training cases and this operation is more inefficient with respect to  PCA. Moreover, if we want to represent the input non-linearly (using curved dimensions) the training of the neural network will become more efficient compared to other methods because its complexity remains linear in the number of training cases.
We can extend the neural network by injecting more hidden layers before and after the original one. The new subnetwork  added before the original hidden layer will be trained to encode the input in the new latent space while the one we added after will decode the data in the original space. As known initializing weights in deep NN can be a non trivial task and this problem can be alleviated performing a pre-training operation or using the same methods by Echo-state networks. One of the first and more famous autoencoder \cite{hinton2006reducing} encoded 784 pixel images in a 30 dimensions space, initializing the weights using Restricted Boltzmann Machines (RBM).

Autoencoders can obviously be employed for entities different than images. One of the first examples was the Latent Semantic Analysis \cite{Letsche:1997}, in which the basic idea was to exploit the PCA to word count vectors to extract similarities between documents. This is a very common task in information retrieval that has been proved to be much better accomplished using autoencoders \cite{autoencoders-lsi}. We can trivially convert documents into bags-of-words and generate the word count vectors. Extracting similarities in the high dimensional space of the words can be a really expensive operation and can be addressed using a much more compact representation using autoencoders for both documents and queries. The most important difference with the previous example is that, in this case, what we want to predict is a probability distribution representing the chance of encountering a specific word in a document, and this affects the model in terms of cost function and optimization objectives.

\subsection{Semantics-aware autoencoders}\label{sec:architecture}
\begin{figure}[t!]
	\includegraphics[width=\linewidth]{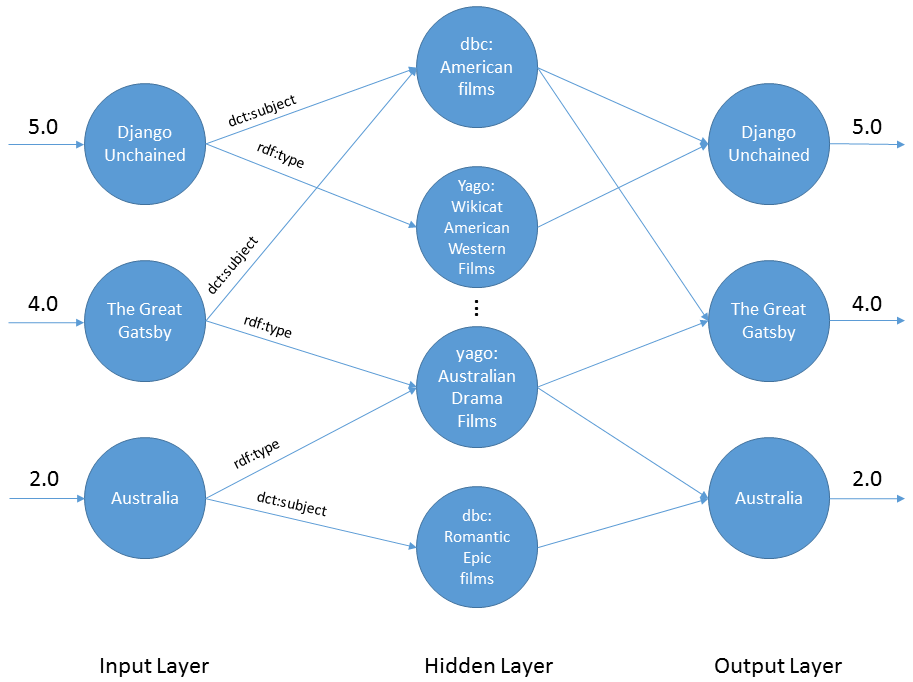}
	\caption{Architecture of a semantic autoencoder.}
	\label{fig:struct}
\end{figure}

Autoencoders, just like other methods for latent representation, are unable to provide an explanation for the latent factors they provide. To address this issue, we propose to give a meaning to connection with the hidden layer and to its neurons by exploiting  semantic information explicitly available in knowledge graphs.  The main idea of the SEM-AUTO approach is to map connections between units from layer \texttt{i} to layer \texttt{i+1}, reflecting the nodes available in a knowledge-based graph (KG) as shown in figure \ref{fig:struct}. In particular, we mapped the autoencoder network topology with the categorical information related to items rated by users. The mapping with KG makes the hidden layer of SEM-AUTO of variable length in the number of units, depending on how much categorical information is available for items rated by a specific user. Items not found in the knowledge graph are just ignored because we cannot retrieve any content description about them. 
Suppose a user $u$ with $n$ rating. Let $m\leq n$ be the number of items rated by $u$ available in the graph and  $C_i = \{c_{i1}, c_{i2}, \dots, c_{ij}\}$ be the set of $j$ categorical nodes associated in the KG to the item $i$. Then, $S = \bigcup_{i=1}^{m} C_{i}$ is the set of features mapped into the hidden layer for the user and the number of hidden units in SEM-AUTO is equal to $\left\vert{S}\right\vert$. Our assumption is that most of the valuable information encoded in a knowledge graph is represented by categorical information. Therefore, meanwhile SEM-AUTO retrieves categorical information of user's rated items from KG, it builds the network on-the-fly by reflecting the graph topology. Once the network is built, the training process for a user takes place feeding the neural network with the item's ratings provided by the user, normalized by \texttt{[0,1]}.

It is worth noticing that the neural network we build is not fully connected. 

In order to make the results of the network consistent during multiple trainings, weights are not initialized randomly, but to a very small value close to zero. We found that the smaller the weights, the better the network convergence with smaller root mean squared error. 
As the nodes in the hidden layer correspond to categories in the knowledge graph, once the model has been trained, the sum of the  weights of edges entering a node represents somehow its worthiness in the definition of a rating. If we consider the nodes associated (connected) to a specific item, their weight may be considered as an initial form of explanation for a given rating.
  
Please note, that such autoencoders do not need bias nodes because these latter are not representative of any semantic data in the graph. Hence, the structure of a generic hidden units looks like the one depicted in Figure \ref{fig:hs}.

\begin{tikzpicture}[
init/.style={
	draw,
	circle,
	inner sep=2pt,
	font=\Huge,
	join = by -latex
},
squa/.style={
	draw,
	inner sep=2pt,
	font=\Large,
	join = by -latex
},
start chain=2,node distance=13mm
]
\node[on chain=2] 
{$w_2$};
\node[on chain=2,init] (sigma) 
{$\displaystyle\Sigma$};
\node[on chain=2,squa,label=above:{\parbox{2cm}{\centering Activation \\ function}}]   
{$f$};
\node[on chain=2,label=above:Output,join=by -latex] 
{$y$};
\begin{scope}[start chain=1]
\node[on chain=1] at (0,1.5cm) 
(w1) {$w_1$};
\end{scope}
\begin{scope}[start chain=3]
\node[on chain=3] at (0,-1.5cm) 
(w3) {$w_3$};
\end{scope}

\draw[-latex] (w1) -- (sigma);
\draw[-latex] (w3) -- (sigma);

\end{tikzpicture}
\captionof{figure}{Structure of units}
\label{fig:hs}
As activation function, in our implementation we used the well known sigmoid function $	\sigma(x) = \frac{1}{1+e^{-x}}$.


%
As a result, once the network converges we have a latent representation of features associated to a user profile together with their weight. However, very interestingly, this time the features also have an explicit meaning as they are in a one to one mapping with elements (nodes) in a knowledge graph.  Our autoencoder is therefore capable to learn for each user the semantics behind her ratings and weight them.

The rationale behind our idea is to catch the side information shared among items rated by a user, so that it gains higher value, tending towards 1, for positively rated items. On the other side, information shared among negatively rated item will tend towards 0. Therefore, our autoencoder finds the best value for those features that approximate the target user ratings by taking into account what are the best features the user is interested in. Features in a user profile are normalized within the interval $[0,1]$ using a standard min-max normalization. 

Given the trained autoencoder, a user profile is then built by considering the features associated to items she rated in the past.

It is worth noticing that in order to train our autoencoder, according to the semantic structure of the data contained in the hidden layer, at least 2 ratings are required for each user, otherwise a constant value for all the features would be spread to approximate the targeted user ratings.

\section{Computing Recommendations}\label{sec:recommendations}
As we said before the weight associated to a feature $f_n$ is the summation of the weights $w_{jn}$ computed in the semantic autoencoder for each edge entering the node representing the feature itself. As we train an auto encoder for each user, we have weights changing depending on the original user profile $P(u) = \{\langle i, r\rangle\}$ with $i$ being an item rated by the user and $r$ its associated rating. More formally, we have
\[
w(f_n,u) = \sum_{j=0}^{j=inndeg(f_n)} w_{jn}
\]
where $inndeg(f_n)$ is the number of edges entering the node representing the feature $f_n$. 
\begin{figure}[t!]
	\includegraphics[width=0.7\linewidth]{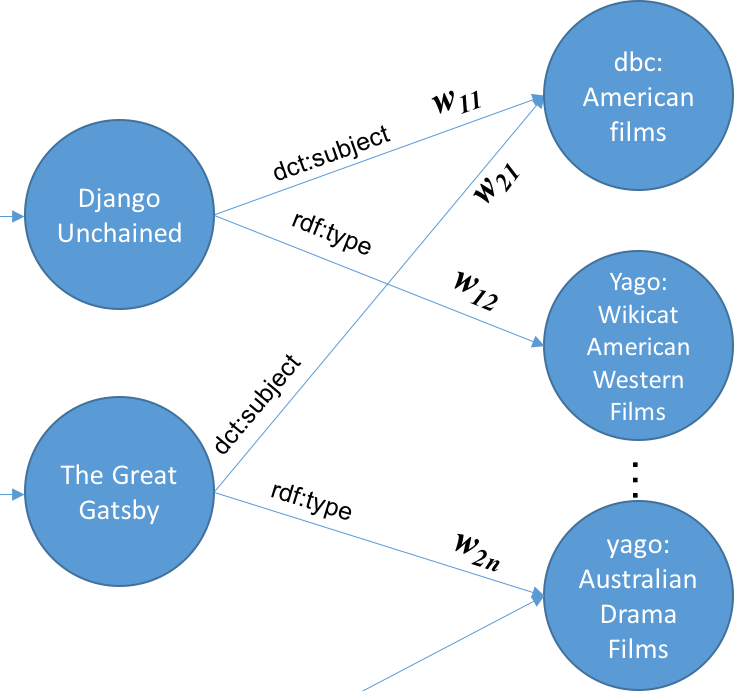}
	\caption{An excerpt of the network in Figure \ref{fig:struct} after the training.}
	\label{fig:trained}
\end{figure}
As an example, if we consider the excerpt of the network in Figure \ref{fig:struct} represented in Figure \ref{fig:trained}, for \texttt{American\_films} we have 
\[
w(\mathtt{American\_Films}, u) = w_{11} + w_{21}
\]
By means of  the weights associated to each feature, we can now move a user profile $P(u)$ from the user$\times$item space to the user$\times$feature one. 
Given $F(i)$ as the set of features composing an item defined as
\[
F(i) = \{f_n ~|~ \text{there is an edge in the autoencoder between } i \text{ and  } f_n\}
\]
we represent the user profile as
\begin{equation}
\hat{P}(u) = \{\langle f_n, w(f_n,u) \rangle ~|~ f_n \in F(i) \text{ with } i \text{ being rated by } u\}
\end{equation}
%
%
%
%
%

Due to the high sparseness of the feature-item matrix, the pure content-based information available in the user profile could not be enough to provide valuable recommendations. Hence, we exploited collaborative information available in the original dataset to further enhance user profiles.

We projected them in a Vector Space Model where each feature is a dimension of the vector space and computed the cosine similarity between users and, for each user we computed the set $K(u)$ containing the $k$ users most similar to $u$. Then, for each feature $f_n' \not\in \hat{P}(u)$ and for each user $u' \in K(u)$ we estimated $w(f_n',u)$ as
\begin{equation}
w(f_n',u) = \frac{\displaystyle \sum_{u' \in K(u)} w(f_n',u')}{k} \label{eq:completion}
\end{equation}
and eventually we added $\langle f_n', w(f_n',u) \rangle$ to $\hat{P}(u)$. 

After this post-processing step, ratings for unknown items $\tilde{i}$ to $u$ can be computed by combining the weights in the user profiles associated to $F(\tilde{i})$. In our implementation we just sum their values.
\begin{equation}
r(\tilde{i},u) = \displaystyle \sum_{f_n \in F(\tilde{i})} w(f_n,u)
\end{equation}


%
%

\section{Experiments}\label{sec:experiments}
In this section, we present the experimental evaluations. We describe the structure of the dataset used in the experiments and the evaluation protocol. In this experimental setup we focused on cold-users with a number of ratings equal to 2, 5 or 10.

\subsection{Dataset}
We conducted the experiment on the Movielens 1M dataset, which is composed by 6040 users and 3952 items. Each user has at least 20 ratings and ratings are made on a 5-star scale.

In our experiments, we referred to the freely available knowledge graph of DBpedia\footnote{\url{https://dbpedia.org}}. In order to map items in Movielens to resources in DBpedia we adopted a freely available mapping\footnote{\url{https://github.com/sisinflab/LODrecsys-datasets}}. The mapping contains 3573 mapped movies of 3952 total movies in the dataset. For each item we extracted categorical information by considering the two RDF predicates:
\begin{description}
	\item[ ] \texttt{http://purl.org/dc/terms/subject}
	\item[ ] \texttt{http://www.w3.org/1999/02/22-rdf-syntax-ns\#type}
\end{description}
The former links to categories as the one available in Wikipedia while the latter is used to classify in a more engineered ontology all the resources available in DBpedia.

\subsection{Evaluation protocol}
Here, we show how we evaluate performances of our methods in recommending items to cold-start users.
We split the dataset using Hold-Out 80/20, ensuring that every user have 80\% of their ratings in the training set and the remaining 20\% in the test set. We look for users in the test set that have at least 10 rates, and we selected them as potential cold user candidates. The protocol presented in the following is inspired by \cite{Ice-Breaking}. We made the candidate users cold by removing their ratings from the training set. We tested our approach with profiles reduced to 2, 5 and 10 ratings.\\

\begin{enumerate}
	\item Setup the cold-start user scenario
	\begin{itemize}
		\item Randomly choose 25\% of users from cold user candidates and put them into set $\mathbb{U}_{c}$
		\item $\forall u\in\mathbb{U}_{c}$ move out their ratings from the training set to $\mathbb{F}_{c}$
	\end{itemize}
	\item Evaluate the cold-start user scenario
	\begin{itemize}
		\item Create an empty set $\mathbb{R}_{c}$
		\item For $ n \in \{ 2, 5, 10 \} $ do
		\begin{itemize}
			\item $\forall u\in\mathbb{U}_{c}$ do:
				\begin{itemize}
					\item randomly pick up $n$ of his ratings from $\mathbb{F}_{c}$ and move them to the training set
				\end{itemize}
			\item Train the model
			\item $\forall u\in\mathbb{U}_{c}$ generate recommendation for all unrated items 
			\item Evaluate recommendations for cold-users only
		\end{itemize}
	\end{itemize}
\end{enumerate}

\subsection{Metrics}
In this work we avoided to use Root Mean Squared Error (RMSE). It is known that it may estimate the same error for top-N items and bottom-N items, without taking into account that an error in top-N items should be more relevant compared to an error for lower ranked items.
For this reason, we chose to use Precision and Recall and nDCG to evaluate the accuracy of our model in cold user scenarios. 

Precision is defined as the proportion of retrieved items that are relevant to the user.

\[
	Precision@N = \frac{|L_{u}(N) \cap TS_{u}^{+}|}{N}
\]
where $L_{u}(N)$ is the recommendation list up to the N-th element and $TS_{u}^{+}$ is the set of relevant test items for $u$. Precision measures the system's ability to reject any non-relevant documents in the retrieved set.

Recall is defined as the proportion of relevant items that are retrieved.

\[
	Recall@N = \frac{|L_{u}(N) \cap TS_{u}^{+}|}{TS_{u}^{+}}
\]

Recall measures the system's ability to find all the relevant documents.

Precision and recall can be combined with each other in the F1 measure computed as the harmonic mean between precision and recall.

\[
F1@N = 2 \cdot \frac{Precision@N \cdot Recall@N}{Precision@N + Recall@N} 
\]

In information retrieval, Discounted cumulative gain (DCG) is a metric of ranking quality that measures the usefulness of a document based on its position in the result list. Recommended results may vary in length depending on the user, therefore is not possibile to compare performance among different users, so the cumulative gain at each position should be normalized across users. Hence, normalized discounted cumulative gain, or nDCG, is computed as:

\[
	nDCG_{u}@N = \frac{1}{IDCG@N} \sum_{p=1}^{N} \frac{2^{r_{up}}-1}{\log_{2}(1+p)}
\]
where $p$ is the position of an item in the recommendation list and  $IDCG@N$ indicates the score obtained by an ideal ranking of $L_{u}(N)$.

Accuracy metrics are a valuable way to evaluate the performance of a recommender system. Nonetheless, it has been argued \cite{Smyth2001} that also diversity should be taken into account when evaluating users' satisfaction. In order to evaluate the diversification power of our approach we also measure ERR-IA\cite{chapelle2009expected}.  
\[
ERR-IA = \sum_{r=1}^{n}{\frac{1}{r}\sum_t{P(t|q)\prod_{i=1}^{r-1}{(1-R_i^t)R_r^t}}}
\]
where $r$ is the position of an item $i$, $t$ is the topic (in our case topics are movie genres as stated in Movielens Datatset ancillary files), $P(t|q)$ is the conditional probability of the topic given the query (user profiles in this case), $R_i$ is the probability of the relevance of the item and  $R_r$ is the probability of the relevance of the list of items from $1$ to $r$. 
With this metric, the contribution of each item in the recommendation list is based on the relevance of documents ranked above it. The discount function then depends also on the relevance of previously ranked documents.

\subsection{Results Discussion}\label{sec:discussion}
\begin{table*}[h!]
	\centering
	\begin{tabular}{l|c|c|c|c|c|c|c|}
		\cline{2-8}
		& \textbf{\#ratings} & \textbf{k} & {\textbf{f1@10}} & {\textbf{precision@10}} & {\textbf{recall@10}} & {\textbf{nDCG@10}} & {\textbf{ERR-IA@10}} \\
		\hline
		\textbf{BPRSLIM} & \multirow{2}{*}{2} &   $-$    & 0.021741632 & 0.032649007 & 0.016297099 & 0.023353576 & 0.018825394 \\
		\cline{1-1}\cline{3-8}
		\textbf{SEM-AUTO} &       & 10    & \textbf{0.023096283} & \textbf{0.033046358} & \textbf{0.017751427} & \textbf{0.028283378} & \textbf{0.02600731} \\
		\hline\hline
		\textbf{BPRSLIM} & \multirow{2}{*}{5} &  $-$     & 0.039078954 & 0.050066225 & 0.032046252 & 0.045158629 & 0.042121369 \\
		\cline{1-1}\cline{3-8}
		\textbf{SEM-AUTO} &       & 100   & 0.038598535 & \textbf{0.054039735} & 0.030020531 & \textbf{0.048623943} & \textbf{0.047124717} \\
		\hline
	\end{tabular}%
	\caption{Experimental Results. \textbf{\#ratings} represents the number of ratings in cold users. \textbf{k} is the number of similar users belonging to $K(u)$ used in Equation (\ref{eq:completion})} \label{tab:results}%
\end{table*}%

\begin{figure}[h!]
	\includegraphics[width=\linewidth]{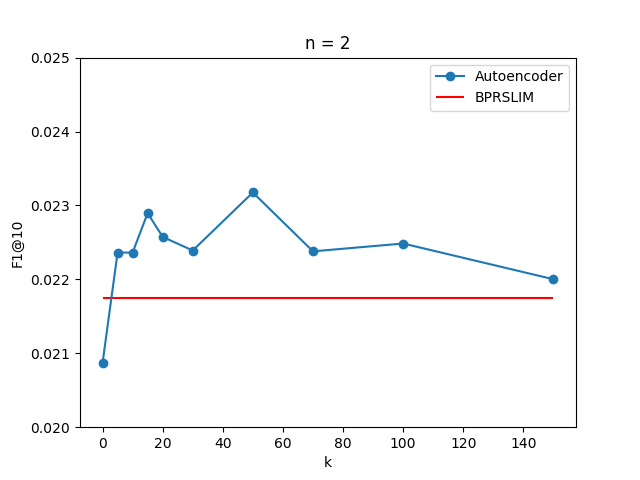}
	\caption{Plot for 2 ratings.}
	\label{fig:plot-n2}
\end{figure}

\begin{figure}[h!]
	\includegraphics[width=\linewidth]{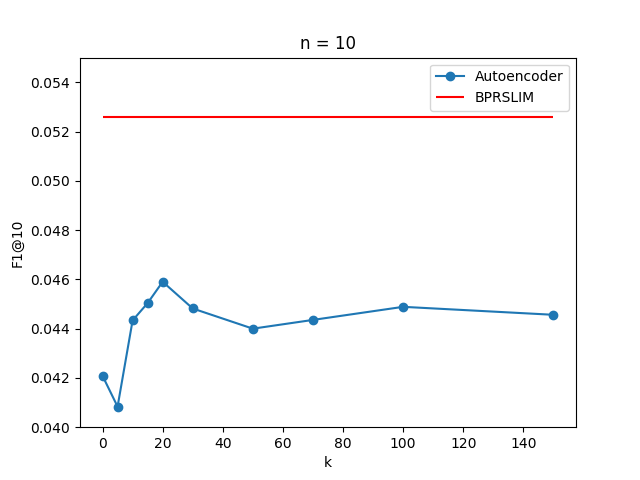}
	\caption{Plot for 10 ratings.}
	\label{fig:plot-n10}
\end{figure}

In our experiments, we compared our approach with the implementation of BPRSLIM \cite{bpr,slim} available in MyMediaLite\footnote{\url{http://mymedialite.net}} \cite{Gantner2011MyMediaLite} as baseline. BPRSLIM is a CF state-of-the-art sparse linear method that leverages the objective function as Bayesian personalized ranking.
In Table \ref{tab:results} we report only those configurations for which our semantic-autoencoder gets the best results compared to BPRSLIM. We can see that for a number of ratings equal to 2 and 5, we outperform BPRSLIM in terms of precision and nDCG. From Table \ref{tab:results} we see that our approach gets much better results also in terms of recall and ERR-IA for very cold users, i.e., with only 2 ratings in the profile. As the number of ratings grows, the collaborative component becomes more relevant and BPRSLIM beats our SEM-AUTO approach. This result is more evident if we compare the plots in Figure \ref{fig:plot-n2} and \ref{fig:plot-n10} reporting the value of F1 for users with 2 and 5 ratings in their profile respectively. In the former, with only 2 ratings,  SEM-AUTO shows a much better behavior than BPRSLIM while in the latter case BPRSLIM is never beaten by SEM-AUTO. It is interesting to note that, depending on the number of ratings in the user profile,the performance in term of accuracy decreases as the number of neighbors increases. Another interesting result here is that increasing the number of ratings we need a higher number of neighbors in $K(u)$ to compute $w(f_n',u)$ in Equation (\ref{eq:completion}) to reach results comparable with BPRSLIM.
As for diversity, in very cold user situations, SEM-AUTO shows to diversify recommendation results better than BPRSLIM.

%

%

\section{Conclusion and Future Work}\label{sec:conclusion}
In this paper, we have presented a novel method to design semantics-aware autoencoders (SEM-AUTO) driven by information encoded in knowledge graphs. As for classical applications of autoencoders to feature selection, we compute a latent representation of items but, in our case, we can attach an explicit semantics to selected features. This allows our system to exploit both the power of deep learning techniques and, at the same time to have a meaningful and understandable representation of the trained model. We used our approach to autoencode user ratings in a recommendation scenario via the DBpedia knowledge graph and proposed a simple algorithm to compute recommendations based on the semantic features we extract with our autoencoder. Experimental results show that even with a simple approach that just sums the weights associated to features we are able to beat state of the art recommendation algorithms for cold user scenarios. 
The preliminary results presented in this paper pave the way to various further investigations. We are currently testing  SEM-AUTO on other knowledge graphs such as Wikidata to see how much our results depend on the underlying semantic data. Moreover, we want to exploit the features extracted from SEM-AUTO as side information in hybrid state of the art approaches to test their representational effectiveness. Finally, having an explicit representation of latent features opens the door to a better user modeling by means of preference based languages such as CP-theories \cite{dinoia-ijcai-15} that can be further exploited to provide meaningful explanations to the user for recommendation results.






\begin{thebibliography}{00}


\ifx \showCODEN    \undefined \def \showCODEN     #1{\unskip}     \fi
\ifx \showDOI      \undefined \def \showDOI       #1{#1}\fi
\ifx \showISBNx    \undefined \def \showISBNx     #1{\unskip}     \fi
\ifx \showISBNxiii \undefined \def \showISBNxiii  #1{\unskip}     \fi
\ifx \showISSN     \undefined \def \showISSN      #1{\unskip}     \fi
\ifx \showLCCN     \undefined \def \showLCCN      #1{\unskip}     \fi
\ifx \shownote     \undefined \def \shownote      #1{#1}          \fi
\ifx \showarticletitle \undefined \def \showarticletitle #1{#1}   \fi
\ifx \showURL      \undefined \def \showURL       {\relax}        \fi
\providecommand\bibfield[2]{#2}
\providecommand\bibinfo[2]{#2}
\providecommand\natexlab[1]{#1}
\providecommand\showeprint[2][]{arXiv:#2}

\bibitem[\protect\citeauthoryear{Auer, Bizer, Kobilarov, Lehmann, Cyganiak, and
  Ives}{Auer et~al\mbox{.}}{2007}]%
        {dbpedia2007}
\bibfield{author}{\bibinfo{person}{S\"{o}ren Auer}, \bibinfo{person}{Christian
  Bizer}, \bibinfo{person}{Georgi Kobilarov}, \bibinfo{person}{Jens Lehmann},
  \bibinfo{person}{Richard Cyganiak}, {and} \bibinfo{person}{Zachary Ives}.}
  \bibinfo{year}{2007}\natexlab{}.
\newblock \showarticletitle{DBpedia: A Nucleus for a Web of Open Data}. In
  \bibinfo{booktitle}{{\em Proceedings of the 6th International The Semantic
  Web and 2Nd Asian Conference on Asian Semantic Web Conference}} {\em
  (\bibinfo{series}{ISWC'07/ASWC'07})}. \bibinfo{publisher}{Springer-Verlag},
  \bibinfo{pages}{722--735}.
\newblock


\bibitem[\protect\citeauthoryear{Chapelle, Metlzer, Zhang, and
  Grinspan}{Chapelle et~al\mbox{.}}{2009}]%
        {chapelle2009expected}
\bibfield{author}{\bibinfo{person}{Olivier Chapelle}, \bibinfo{person}{Donald
  Metlzer}, \bibinfo{person}{Ya Zhang}, {and} \bibinfo{person}{Pierre
  Grinspan}.} \bibinfo{year}{2009}\natexlab{}.
\newblock \showarticletitle{Expected reciprocal rank for graded relevance}. In
  \bibinfo{booktitle}{{\em Proceedings of the 18th ACM conference on
  Information and knowledge management}}. ACM, \bibinfo{pages}{621--630}.
\newblock


\bibitem[\protect\citeauthoryear{Covington, Adams, and Sargin}{Covington
  et~al\mbox{.}}{2016}]%
        {covington2016deep}
\bibfield{author}{\bibinfo{person}{Paul Covington}, \bibinfo{person}{Jay
  Adams}, {and} \bibinfo{person}{Emre Sargin}.}
  \bibinfo{year}{2016}\natexlab{}.
\newblock \showarticletitle{Deep neural networks for youtube recommendations}.
  In \bibinfo{booktitle}{{\em Proceedings of the 10th ACM Conference on
  Recommender Systems}}. ACM, \bibinfo{pages}{191--198}.
\newblock


\bibitem[\protect\citeauthoryear{{Di Noia}, Lukasiewicz, Mart{\'{\i}}nez,
  Simari, and Tifrea{-}Marciuska}{{Di Noia} et~al\mbox{.}}{2015}]%
        {dinoia-ijcai-15}
\bibfield{author}{\bibinfo{person}{Tommaso {Di Noia}}, \bibinfo{person}{Thomas
  Lukasiewicz}, \bibinfo{person}{Maria~Vanina Mart{\'{\i}}nez},
  \bibinfo{person}{Gerardo~I. Simari}, {and} \bibinfo{person}{Oana
  Tifrea{-}Marciuska}.} \bibinfo{year}{2015}\natexlab{}.
\newblock \showarticletitle{Combining Existential Rules with the Power of
  CP-Theories}. In \bibinfo{booktitle}{{\em Proceedings of the Twenty-Fourth
  International Joint Conference on Artificial Intelligence, {IJCAI} 2015,
  Buenos Aires, Argentina, July 25-31, 2015}}. \bibinfo{pages}{2918--2925}.
\newblock


\bibitem[\protect\citeauthoryear{Dong, Yu, Wu, Sun, Yuan, and Zhang}{Dong
  et~al\mbox{.}}{2017}]%
        {AAAI1714676}
\bibfield{author}{\bibinfo{person}{Xin Dong}, \bibinfo{person}{Lei Yu},
  \bibinfo{person}{Zhonghuo Wu}, \bibinfo{person}{Yuxia Sun},
  \bibinfo{person}{Lingfeng Yuan}, {and} \bibinfo{person}{Fangxi Zhang}.}
  \bibinfo{year}{2017}\natexlab{}.
\newblock \bibinfo{title}{A Hybrid Collaborative Filtering Model with Deep
  Structure for Recommender Systems}.
\newblock   (\bibinfo{year}{2017}).
\newblock
\showURL{%
\url{https://aaai.org/ocs/index.php/AAAI/AAAI17/paper/view/14676}}


\bibitem[\protect\citeauthoryear{Elkahky, Song, and He}{Elkahky
  et~al\mbox{.}}{2015}]%
        {Elkahky:2015:MDL:2736277.2741667}
\bibfield{author}{\bibinfo{person}{Ali~Mamdouh Elkahky}, \bibinfo{person}{Yang
  Song}, {and} \bibinfo{person}{Xiaodong He}.} \bibinfo{year}{2015}\natexlab{}.
\newblock \showarticletitle{A Multi-View Deep Learning Approach for Cross
  Domain User Modeling in Recommendation Systems}. In \bibinfo{booktitle}{{\em
  Proceedings of the 24th International Conference on World Wide Web}} {\em
  (\bibinfo{series}{WWW '15})}. \bibinfo{publisher}{International World Wide
  Web Conferences Steering Committee}, \bibinfo{address}{Republic and Canton of
  Geneva, Switzerland}, \bibinfo{pages}{278--288}.
\newblock
\showISBNx{978-1-4503-3469-3}
\showDOI{%
\url{https://doi.org/10.1145/2736277.2741667}}


\bibitem[\protect\citeauthoryear{Fern\'{a}ndez-Tob\'{\i}as, Cantador,
  Kaminskas, and Ricci}{Fern\'{a}ndez-Tob\'{\i}as et~al\mbox{.}}{2011}]%
        {Fernandez-Tobias2011}
\bibfield{author}{\bibinfo{person}{Ignacio Fern\'{a}ndez-Tob\'{\i}as},
  \bibinfo{person}{Iv\'{a}n Cantador}, \bibinfo{person}{Marius Kaminskas},
  {and} \bibinfo{person}{Francesco Ricci}.} \bibinfo{year}{2011}\natexlab{}.
\newblock \showarticletitle{A generic semantic-based framework for cross-domain
  recommendation}. In \bibinfo{booktitle}{{\em Proceedings of the 2nd
  International Workshop on Information Heterogeneity and Fusion in Recommender
  Systems}} {\em (\bibinfo{series}{HetRec '11})}.
\newblock


\bibitem[\protect\citeauthoryear{Fern\'{a}ndez-Tob\'{\i}as, Tomeo, Cantador,
  Di~Noia, and Di~Sciascio}{Fern\'{a}ndez-Tob\'{\i}as et~al\mbox{.}}{2016}]%
        {tomeoetal2016}
\bibfield{author}{\bibinfo{person}{Ignacio Fern\'{a}ndez-Tob\'{\i}as},
  \bibinfo{person}{Paolo Tomeo}, \bibinfo{person}{Iv\'{a}n Cantador},
  \bibinfo{person}{Tommaso Di~Noia}, {and} \bibinfo{person}{Eugenio
  Di~Sciascio}.} \bibinfo{year}{2016}\natexlab{}.
\newblock \showarticletitle{Accuracy and Diversity in Cross-domain
  Recommendations for Cold-start Users with Positive-only Feedback}. In
  \bibinfo{booktitle}{{\em Proceedings of the 10th ACM Conference on
  Recommender Systems}} {\em (\bibinfo{series}{RecSys '16})}.
  \bibinfo{publisher}{ACM}.
\newblock


\bibitem[\protect\citeauthoryear{Gantner, Rendle, Freudenthaler, and
  Schmidt-Thieme}{Gantner et~al\mbox{.}}{2011}]%
        {Gantner2011MyMediaLite}
\bibfield{author}{\bibinfo{person}{Zeno Gantner}, \bibinfo{person}{Steffen
  Rendle}, \bibinfo{person}{Christoph Freudenthaler}, {and}
  \bibinfo{person}{Lars Schmidt-Thieme}.} \bibinfo{year}{2011}\natexlab{}.
\newblock \showarticletitle{{MyMediaLite}: A Free Recommender System Library}.
  In \bibinfo{booktitle}{{\em 5th ACM International Conference on Recommender
  Systems (RecSys 2011)}}.
\newblock


\bibitem[\protect\citeauthoryear{Heitmann and Hayes}{Heitmann and
  Hayes}{2010}]%
        {Heitmann:2010:firstLODrecsys}
\bibfield{author}{\bibinfo{person}{Benjamin Heitmann} {and}
  \bibinfo{person}{Conor Hayes}.} \bibinfo{year}{2010}\natexlab{}.
\newblock \showarticletitle{C.: Using linked data to build open, collaborative
  recommender systems}. In \bibinfo{booktitle}{{\em In: AAAI Spring Symposium:
  Linked Data Meets Artificial Intelligence’. (2010}}.
\newblock


\bibitem[\protect\citeauthoryear{Hinton and {Salakhutdinov}}{Hinton and
  {Salakhutdinov}}{2006}]%
        {autoencoders-lsi}
\bibfield{author}{\bibinfo{person}{Geoffrey~E. Hinton} {and}
  \bibinfo{person}{Ruslan {Salakhutdinov}}.} \bibinfo{year}{2006}\natexlab{}.
\newblock \showarticletitle{{Reducing the Dimensionality of Data with Neural
  Networks}}.
\newblock \bibinfo{journal}{{\em Science\/}}  \bibinfo{volume}{313}
  (\bibinfo{date}{July} \bibinfo{year}{2006}), \bibinfo{pages}{504--507}.
\newblock
\showDOI{%
\url{https://doi.org/10.1126/science.1127647}}


\bibitem[\protect\citeauthoryear{Hinton and Salakhutdinov}{Hinton and
  Salakhutdinov}{2006}]%
        {hinton2006reducing}
\bibfield{author}{\bibinfo{person}{Geoffrey~E Hinton} {and}
  \bibinfo{person}{Ruslan~R Salakhutdinov}.} \bibinfo{year}{2006}\natexlab{}.
\newblock \showarticletitle{Reducing the dimensionality of data with neural
  networks}.
\newblock \bibinfo{journal}{{\em science\/}} \bibinfo{volume}{313},
  \bibinfo{number}{5786} (\bibinfo{year}{2006}), \bibinfo{pages}{504--507}.
\newblock


\bibitem[\protect\citeauthoryear{Khrouf and Troncy}{Khrouf and Troncy}{2013}]%
        {Khrouf2013}
\bibfield{author}{\bibinfo{person}{Houda Khrouf} {and}
  \bibinfo{person}{Rapha\"{e}l Troncy}.} \bibinfo{year}{2013}\natexlab{}.
\newblock \showarticletitle{Hybrid Event Recommendation Using Linked Data and
  User Diversity}. In \bibinfo{booktitle}{{\em Proceedings of the 7th ACM
  Conference on Recommender Systems}} {\em (\bibinfo{series}{RecSys '13})}.
  \bibinfo{publisher}{ACM}, \bibinfo{address}{New York, NY, USA},
  \bibinfo{pages}{185--192}.
\newblock
\showISBNx{978-1-4503-2409-0}
\showDOI{%
\url{https://doi.org/10.1145/2507157.2507171}}


\bibitem[\protect\citeauthoryear{Kingma and Welling}{Kingma and
  Welling}{2013}]%
        {kingma2013auto}
\bibfield{author}{\bibinfo{person}{Diederik~P Kingma} {and}
  \bibinfo{person}{Max Welling}.} \bibinfo{year}{2013}\natexlab{}.
\newblock \showarticletitle{Auto-encoding variational bayes}.
\newblock \bibinfo{journal}{{\em arXiv preprint arXiv:1312.6114\/}}
  (\bibinfo{year}{2013}).
\newblock


\bibitem[\protect\citeauthoryear{Letsche and Berry}{Letsche and Berry}{1997}]%
        {Letsche:1997}
\bibfield{author}{\bibinfo{person}{Todd~A. Letsche} {and}
  \bibinfo{person}{Michael~W. Berry}.} \bibinfo{year}{1997}\natexlab{}.
\newblock \showarticletitle{Large-scale Information Retrieval with Latent
  Semantic Indexing}.
\newblock \bibinfo{journal}{{\em Inf. Sci.\/}} \bibinfo{volume}{100},
  \bibinfo{number}{1-4} (\bibinfo{year}{1997}), \bibinfo{pages}{105--137}.
\newblock
\showDOI{%
\url{https://doi.org/10.1016/S0020-0255(97)00044-3}}


\bibitem[\protect\citeauthoryear{Meymandpour and Davis}{Meymandpour and
  Davis}{2015}]%
        {Meymandpour:15}
\bibfield{author}{\bibinfo{person}{Rouzbeh Meymandpour} {and}
  \bibinfo{person}{Joseph~G Davis}.} \bibinfo{year}{2015}\natexlab{}.
\newblock \showarticletitle{Enhancing {R}ecommender {S}ystems Using {L}inked
  {O}pen {D}ata-Based Semantic Analysis of Items}. In \bibinfo{booktitle}{{\em
  Proceedings of the 3rd Australasian Web Conference (AWC 2015)}},
  Vol.~\bibinfo{volume}{27}. \bibinfo{pages}{11--17}.
\newblock


\bibitem[\protect\citeauthoryear{Musto, Lops, Basile, de~Gemmis, and
  Semeraro}{Musto et~al\mbox{.}}{2016}]%
        {Musto-UMAP16-LOD}
\bibfield{author}{\bibinfo{person}{Cataldo Musto}, \bibinfo{person}{Pasquale
  Lops}, \bibinfo{person}{Pierpaolo Basile}, \bibinfo{person}{Marco de Gemmis},
  {and} \bibinfo{person}{Giovanni Semeraro}.} \bibinfo{year}{2016}\natexlab{}.
\newblock \showarticletitle{Semantics-aware Graph-based Recommender Systems
  Exploiting Linked Open Data}.
  \bibinfo{howpublished}{http://doi.acm.org/10.1145/2930238.2930249}. In
  \bibinfo{booktitle}{{\em Proceedings of the 2016 Conference on User Modeling
  Adaptation and Personalization, UMAP 2016, Halifax, NS, Canada, July 13 - 17,
  2016}}, \bibfield{editor}{\bibinfo{person}{Julita Vassileva},
  \bibinfo{person}{James Blustein}, \bibinfo{person}{Lora Aroyo}, {and}
  \bibinfo{person}{Sidney K.~D. Mello}} (Eds.). \bibinfo{publisher}{ACM},
  \bibinfo{pages}{229--237}.
\newblock
\showDOI{%
\url{https://doi.org/10.1145/2930238.2930249}}


\bibitem[\protect\citeauthoryear{Musto, Semeraro, Lops, and de~Gemmis}{Musto
  et~al\mbox{.}}{2014}]%
        {MustoSLG14}
\bibfield{author}{\bibinfo{person}{Cataldo Musto}, \bibinfo{person}{Giovanni
  Semeraro}, \bibinfo{person}{Pasquale Lops}, {and} \bibinfo{person}{Marco de
  Gemmis}.} \bibinfo{year}{2014}\natexlab{}.
\newblock \showarticletitle{Combining Distributional Semantics and Entity
  Linking for Context-Aware Content-Based Recommendation}. In
  \bibinfo{booktitle}{{\em User Modeling, Adaptation, and Personalization -
  22nd International Conference, {UMAP} 2014}}.
\newblock


\bibitem[\protect\citeauthoryear{Ning and Karypis}{Ning and Karypis}{2011}]%
        {slim}
\bibfield{author}{\bibinfo{person}{Xia Ning} {and} \bibinfo{person}{George
  Karypis}.} \bibinfo{year}{2011}\natexlab{}.
\newblock \showarticletitle{SLIM: Sparse Linear Methods for Top-N Recommender
  Systems}. In \bibinfo{booktitle}{{\em Proceedings of the 2011 IEEE 11th
  International Conference on Data Mining}} {\em (\bibinfo{series}{ICDM '11})}.
  \bibinfo{publisher}{IEEE Computer Society}, \bibinfo{pages}{497--506}.
\newblock


\bibitem[\protect\citeauthoryear{Ning and Karypis}{Ning and Karypis}{2012}]%
        {Ning12}
\bibfield{author}{\bibinfo{person}{Xia Ning} {and} \bibinfo{person}{George
  Karypis}.} \bibinfo{year}{2012}\natexlab{}.
\newblock \showarticletitle{Sparse linear methods with side information for
  top-n recommendations}. In \bibinfo{booktitle}{{\em Proceedings of the sixth
  ACM conference on Recommender systems}} {\em (\bibinfo{series}{RecSys '12})}.
  \bibinfo{publisher}{ACM}, \bibinfo{address}{New York, NY, USA},
  \bibinfo{pages}{155--162}.
\newblock
\showISBNx{978-1-4503-1270-7}


\bibitem[\protect\citeauthoryear{Noia, Ostuni, Tomeo, and Sciascio}{Noia
  et~al\mbox{.}}{2016}]%
        {DiNoia:2016:SPrank}
\bibfield{author}{\bibinfo{person}{Tommaso~Di Noia},
  \bibinfo{person}{Vito~Claudio Ostuni}, \bibinfo{person}{Paolo Tomeo}, {and}
  \bibinfo{person}{Eugenio~Di Sciascio}.} \bibinfo{year}{2016}\natexlab{}.
\newblock \showarticletitle{SPrank: Semantic Path-Based Ranking for Top-N
  Recommendations Using Linked Open Data}.
\newblock \bibinfo{journal}{{\em ACM Trans. Intell. Syst. Technol.\/}}
  \bibinfo{volume}{8}, \bibinfo{number}{1} (\bibinfo{date}{Sept.}
  \bibinfo{year}{2016}).
\newblock


\bibitem[\protect\citeauthoryear{Ostuni, Di~Noia, Di~Sciascio, and
  Mirizzi}{Ostuni et~al\mbox{.}}{2013}]%
        {Ostuni2013}
\bibfield{author}{\bibinfo{person}{Vito~Claudio Ostuni},
  \bibinfo{person}{Tommaso Di~Noia}, \bibinfo{person}{Eugenio Di~Sciascio},
  {and} \bibinfo{person}{Roberto Mirizzi}.} \bibinfo{year}{2013}\natexlab{}.
\newblock \showarticletitle{Top-N Recommendations from Implicit Feedback
  Leveraging Linked Open Data}. In \bibinfo{booktitle}{{\em Proceedings of the
  7th ACM Conference on Recommender Systems}} {\em (\bibinfo{series}{RecSys
  '13})}. \bibinfo{publisher}{ACM}, \bibinfo{address}{New York, NY, USA},
  \bibinfo{pages}{85--92}.
\newblock
\showISBNx{978-1-4503-2409-0}
\showDOI{%
\url{https://doi.org/10.1145/2507157.2507172}}


\bibitem[\protect\citeauthoryear{Passant}{Passant}{2010}]%
        {Passant2010}
\bibfield{author}{\bibinfo{person}{Alexandre Passant}.}
  \bibinfo{year}{2010}\natexlab{}.
\newblock \bibinfo{booktitle}{{\em dbrec --- Music Recommendations Using
  DBpedia}}.
\newblock


\bibitem[\protect\citeauthoryear{Ragone, Tomeo, Magarelli, Di~Noia, Palmonari,
  Maurino, and Di~Sciascio}{Ragone et~al\mbox{.}}{2017}]%
        {Ragone-2017}
\bibfield{author}{\bibinfo{person}{Azzurra Ragone}, \bibinfo{person}{Paolo
  Tomeo}, \bibinfo{person}{Corrado Magarelli}, \bibinfo{person}{Tommaso
  Di~Noia}, \bibinfo{person}{Matteo Palmonari}, \bibinfo{person}{Andrea
  Maurino}, {and} \bibinfo{person}{Eugenio Di~Sciascio}.}
  \bibinfo{year}{2017}\natexlab{}.
\newblock \showarticletitle{Schema-summarization in Linked-data-based Feature
  Selection for Recommender Systems}. In \bibinfo{booktitle}{{\em Proceedings
  of the Symposium on Applied Computing}} {\em (\bibinfo{series}{SAC '17})}.
  \bibinfo{publisher}{ACM}, \bibinfo{pages}{330--335}.
\newblock
\showDOI{%
\url{https://doi.org/10.1145/3019612.3019837}}


\bibitem[\protect\citeauthoryear{Rendle, Freudenthaler, Gantner, and
  Schmidt-Thieme}{Rendle et~al\mbox{.}}{2009}]%
        {bpr}
\bibfield{author}{\bibinfo{person}{Steffen Rendle}, \bibinfo{person}{Christoph
  Freudenthaler}, \bibinfo{person}{Zeno Gantner}, {and} \bibinfo{person}{Lars
  Schmidt-Thieme}.} \bibinfo{year}{2009}\natexlab{}.
\newblock \showarticletitle{BPR: Bayesian Personalized Ranking from Implicit
  Feedback}. In \bibinfo{booktitle}{{\em Proceedings of the Twenty-Fifth
  Conference on Uncertainty in Artificial Intelligence}} {\em
  (\bibinfo{series}{UAI '09})}. \bibinfo{publisher}{AUAI Press},
  \bibinfo{pages}{452--461}.
\newblock


\bibitem[\protect\citeauthoryear{Sedhain, Menon, Sanner, and Xie}{Sedhain
  et~al\mbox{.}}{2015}]%
        {Sedhain:2015:AAM:2740908.2742726}
\bibfield{author}{\bibinfo{person}{Suvash Sedhain},
  \bibinfo{person}{Aditya~Krishna Menon}, \bibinfo{person}{Scott Sanner}, {and}
  \bibinfo{person}{Lexing Xie}.} \bibinfo{year}{2015}\natexlab{}.
\newblock \showarticletitle{AutoRec: Autoencoders Meet Collaborative
  Filtering}. In \bibinfo{booktitle}{{\em Proceedings of the 24th International
  Conference on World Wide Web}} {\em (\bibinfo{series}{WWW '15 Companion})}.
  \bibinfo{publisher}{ACM}, \bibinfo{address}{New York, NY, USA},
  \bibinfo{pages}{111--112}.
\newblock
\showISBNx{978-1-4503-3473-0}
\showDOI{%
\url{https://doi.org/10.1145/2740908.2742726}}


\bibitem[\protect\citeauthoryear{Smyth and McClave}{Smyth and McClave}{2001}]%
        {Smyth2001}
\bibfield{author}{\bibinfo{person}{Barry Smyth} {and} \bibinfo{person}{Paul
  McClave}.} \bibinfo{year}{2001}\natexlab{}.
\newblock \showarticletitle{Similarity vs. Diversity}. In
  \bibinfo{booktitle}{{\em Proceedings of the 4th International Conference on
  Case-Based Reasoning: Case-Based Reasoning Research and Development}} {\em
  (\bibinfo{series}{ICCBR '01})}. \bibinfo{publisher}{Springer-Verlag},
  \bibinfo{address}{London, UK, UK}, \bibinfo{pages}{347--361}.
\newblock
\showISBNx{3-540-42358-3}
\showURL{%
\url{http://dl.acm.org/citation.cfm?id=646268.758890}}


\bibitem[\protect\citeauthoryear{Strub, Gaudel, and Mary}{Strub
  et~al\mbox{.}}{2016}]%
        {Strub:2016:HRS:2988450.2988456}
\bibfield{author}{\bibinfo{person}{Florian Strub}, \bibinfo{person}{Romaric
  Gaudel}, {and} \bibinfo{person}{J{\'e}r{\'e}mie Mary}.}
  \bibinfo{year}{2016}\natexlab{}.
\newblock \showarticletitle{Hybrid Recommender System Based on Autoencoders}.
  In \bibinfo{booktitle}{{\em Proceedings of the 1st Workshop on Deep Learning
  for Recommender Systems}} {\em (\bibinfo{series}{DLRS 2016})}.
  \bibinfo{publisher}{ACM}, \bibinfo{address}{New York, NY, USA},
  \bibinfo{pages}{11--16}.
\newblock
\showISBNx{978-1-4503-4795-2}
\showDOI{%
\url{https://doi.org/10.1145/2988450.2988456}}


\bibitem[\protect\citeauthoryear{Vincent, Larochelle, Bengio, and
  Manzagol}{Vincent et~al\mbox{.}}{2008}]%
        {vincent2008extracting}
\bibfield{author}{\bibinfo{person}{Pascal Vincent}, \bibinfo{person}{Hugo
  Larochelle}, \bibinfo{person}{Yoshua Bengio}, {and}
  \bibinfo{person}{Pierre-Antoine Manzagol}.} \bibinfo{year}{2008}\natexlab{}.
\newblock \showarticletitle{Extracting and composing robust features with
  denoising autoencoders}. In \bibinfo{booktitle}{{\em Proceedings of the 25th
  international conference on Machine learning}}. ACM,
  \bibinfo{pages}{1096--1103}.
\newblock


\bibitem[\protect\citeauthoryear{Vrande\v{c}i\'{c} and
  Kr\"{o}tzsch}{Vrande\v{c}i\'{c} and Kr\"{o}tzsch}{2014}]%
        {wikidata2014}
\bibfield{author}{\bibinfo{person}{Denny Vrande\v{c}i\'{c}} {and}
  \bibinfo{person}{Markus Kr\"{o}tzsch}.} \bibinfo{year}{2014}\natexlab{}.
\newblock \showarticletitle{Wikidata: A Free Collaborative Knowledgebase}.
\newblock \bibinfo{journal}{{\em Commun. ACM\/}} \bibinfo{volume}{57},
  \bibinfo{number}{10} (\bibinfo{year}{2014}), \bibinfo{pages}{78--85}.
\newblock
\showDOI{%
\url{https://doi.org/10.1145/2629489}}


\bibitem[\protect\citeauthoryear{Vuurens, Larson, and de~Vries}{Vuurens
  et~al\mbox{.}}{2016}]%
        {Vuurens:2016:EDS:2988450.2988457}
\bibfield{author}{\bibinfo{person}{Jeroen B.~P. Vuurens},
  \bibinfo{person}{Martha Larson}, {and} \bibinfo{person}{Arjen~P. de Vries}.}
  \bibinfo{year}{2016}\natexlab{}.
\newblock \showarticletitle{Exploring Deep Space: Learning Personalized Ranking
  in a Semantic Space}. In \bibinfo{booktitle}{{\em Proceedings of the 1st
  Workshop on Deep Learning for Recommender Systems}} {\em
  (\bibinfo{series}{DLRS 2016})}. \bibinfo{publisher}{ACM},
  \bibinfo{address}{New York, NY, USA}, \bibinfo{pages}{23--28}.
\newblock
\showISBNx{978-1-4503-4795-2}
\showDOI{%
\url{https://doi.org/10.1145/2988450.2988457}}


\bibitem[\protect\citeauthoryear{Wang, Wang, and Yeung}{Wang
  et~al\mbox{.}}{2015}]%
        {Wang:2015:CDL:2783258.2783273}
\bibfield{author}{\bibinfo{person}{Hao Wang}, \bibinfo{person}{Naiyan Wang},
  {and} \bibinfo{person}{Dit-Yan Yeung}.} \bibinfo{year}{2015}\natexlab{}.
\newblock \showarticletitle{Collaborative Deep Learning for Recommender
  Systems}. In \bibinfo{booktitle}{{\em Proceedings of the 21th ACM SIGKDD
  International Conference on Knowledge Discovery and Data Mining}} {\em
  (\bibinfo{series}{KDD '15})}. \bibinfo{publisher}{ACM}, \bibinfo{address}{New
  York, NY, USA}, \bibinfo{pages}{1235--1244}.
\newblock
\showISBNx{978-1-4503-3664-2}
\showDOI{%
\url{https://doi.org/10.1145/2783258.2783273}}


\bibitem[\protect\citeauthoryear{Wu, DuBois, Zheng, and Ester}{Wu
  et~al\mbox{.}}{2016}]%
        {Wu2016}
\bibfield{author}{\bibinfo{person}{Yao Wu}, \bibinfo{person}{Christopher
  DuBois}, \bibinfo{person}{Alice~X. Zheng}, {and} \bibinfo{person}{Martin
  Ester}.} \bibinfo{year}{2016}\natexlab{}.
\newblock \showarticletitle{Collaborative Denoising Auto-Encoders for Top-N
  Recommender Systems}. In \bibinfo{booktitle}{{\em Proceedings of the Ninth
  ACM International Conference on Web Search and Data Mining}} {\em
  (\bibinfo{series}{WSDM '16})}. \bibinfo{publisher}{ACM},
  \bibinfo{address}{New York, NY, USA}, \bibinfo{pages}{153--162}.
\newblock
\showISBNx{978-1-4503-3716-8}
\showDOI{%
\url{https://doi.org/10.1145/2835776.2835837}}


\bibitem[\protect\citeauthoryear{Xu, Yao, Tong, Tao, and Lu}{Xu
  et~al\mbox{.}}{2015}]%
        {Ice-Breaking}
\bibfield{author}{\bibinfo{person}{Jingwei Xu}, \bibinfo{person}{Yuan Yao},
  \bibinfo{person}{Hanghang Tong}, \bibinfo{person}{Xianping Tao}, {and}
  \bibinfo{person}{Jian Lu}.} \bibinfo{year}{2015}\natexlab{}.
\newblock \bibinfo{booktitle}{{\em Ice-Breaking: Mitigating cold-start
  recommendation problem by rating comparison}}.
\newblock \bibinfo{pages}{3981--3987}.
\newblock


\end{thebibliography}
\end{document}